%% file: main.tex
\documentclass[runningheads]{llncs}

\usepackage{malte}

\hypersetup{hidelinks}

\usepackage[%
    backend=biber,
    style=numeric-comp,
    dateabbrev=false,
    maxbibnames=99,
    sortcites=true,
    hyperref=auto,
    doi=false,
    isbn=false,
    eprint=false]{biblatex}
\usepackage{custom}

\newtoggle{blind}
\togglefalse{blind}

\newtoggle{draft}
\togglefalse{draft}

\iftoggle{draft}{%
    \newcommand{\anote}[1]{{\color{red}[ {{#1}} -- AN]}}%
    \newcommand{\response}[1]{{\color{sngreen!80!black}$\rightarrow$ \textit{#1}}}%
    \newcommand{\malte}[1]{{\color{snred!80!black}\textit{#1}}}
    \newcommand{\todo}[1]{{\color{snblue}TODO: \textit{#1}}}
    
}{%
    \newcommand{\anote}[1]{}%
    \newcommand{\response}[1]{}%
    \newcommand{\malte}[1]{}%
    \newcommand{\todo}[1]{}%
}%

\title{Resurrecting Address Clustering in Bitcoin}

\iftoggle{blind}{
\author{}
\institute{}
}{
\author{Malte Möser \and
Arvind Narayanan}
\authorrunning{M. Möser \and A. Narayanan}
\institute{Princeton University\\\email{mail@maltemoeser.de, arvindn@cs.princeton.edu}}
}

\begin{document}

\maketitle

\begin{abstract}
Blockchain analysis is essential for understanding how cryptocurrencies like Bitcoin are used in practice, and address clustering is a cornerstone of blockchain analysis. However, current techniques rely on heuristics that have not been rigorously evaluated or optimized. In this paper, we tackle several challenges of change address identification and clustering. First, we build a ground truth set of transactions with known change from the Bitcoin blockchain that can be used to validate the efficacy of individual change address detection heuristics. Equipped with this data set, we develop new techniques to predict change outputs with low false positive rates. After applying our prediction model to the Bitcoin blockchain, we analyze the resulting clustering and develop ways to detect and prevent cluster collapse. Finally, we assess the impact our enhanced clustering has on two exemplary applications.
\end{abstract}

\input{paper}

\iftoggle{blind}{%
}{%
\section*{Acknowledgements}

We thank Rainer Böhme and Kevin Lee for their feedback on an earlier draft of this paper. This work is supported by NSF Award CNS-1651938 and a grant from the Ripple University Blockchain Research Initiative.
}

\printbibliography

\newpage
\appendix
\input{appendix}

\end{document}

%% file: paper.tex
\section{Introduction}

Blockchain analysis techniques are essential for understanding how cryptocurrencies like Bitcoin are used in practice.
A major challenge in analyzing blockchains is grouping transactions belonging to the same user.
Users can create an unlimited amount of addresses to receive and send coins.
As a result, their activity is often split among a multitude of such addresses.
\textit{Address clustering heuristics} aim to identify addresses under an individual user's control based on assumptions about how wallets create transactions.
As the term \textit{heuristic} suggests, address clustering today is more intuitive than rigorous; our overarching goal in this paper is to elevate it to a science.

There are at least four applications for which accurate address clustering is important.
First, a law enforcement agency may be interested in evaluating the transactions of a specific entity. %
They may supplement their own investigation with a set of reliable heuristics to identify relevant transactions.
Second, and conversely, the ability to accurately determine a user's transactions directly impacts their privacy.
This tension between law enforcement needs and everyday users' privacy is inherent to cryptocurrencies due to their transparency and pseudonymity.
Advocates from one side push for greater privacy and from the other side for stronger regulation.
To better understand this tug-of-war, it is important to quantify how reliable change address heuristics are in practice.
Third, accurate grouping of transaction activity is important for aggregate analyses such as studying economic activity over time.
This usually requires a full clustering of all addresses on the blockchain. %
Finally, the unique challenges of address clustering may be interesting for researchers outside  of cryptocurrencies.
For example, it may pose as an application domain for machine learning models and could be used as a benchmarking application.%

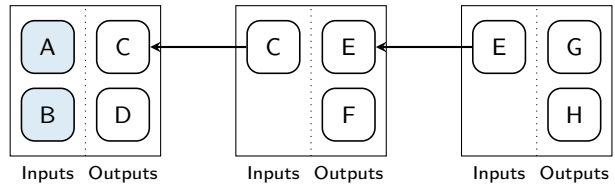
\begin{figure}[!t]
    \centering
    \begin{tikzpicture}[
        inout/.style={
            rectangle,
            semithick,
            draw,
            rounded corners=5pt,
            minimum width=.7cm,
            minimum height=.7cm,
            font=\sffamily
        },
        lbl/.style={
            font=\scriptsize\sffamily
        }
    ]
        \draw[] (0,0) rectangle ++(2,2);
        \draw[dotted] (1,0) -- ++ (0,2);
        \node[inout, fill=snblue!15] at (.5, 1.45) {A};
        \node[inout, fill=snblue!15] at (.5, .55) {B};
        \node[inout] at (1.5, 1.45) (output) {C};
        \node[inout] at (1.5, .55) {D};
        
        \node[lbl] at (.5, -.25) {Inputs};
        \node[lbl] at (1.5, -.25) {Outputs};
        
        \draw[] (3,0) rectangle ++(2,2);
        \draw[dotted] (4,0) -- ++(0,2);
        \node[inout] at (3.5, 1.45) (input) {C};
        \node[inout] at (4.5, 1.45) (output2) {E};
        \node[inout] at (4.5, .55) {F};
        
        \node[lbl] at (3.5, -.25) {Inputs};
        \node[lbl] at (4.5, -.25) {Outputs};
        
        \draw[>=stealth,->, thick] (input) -- (output);
        
        \draw[] (6,0) rectangle ++(2,2);
        \draw[dotted] (7,0) -- ++(0,2);
        \node[inout] at (6.5, 1.45) (input2) {E};
        \node[inout] at (7.5, 1.45) {G};
        \node[inout] at (7.5, .55) {H};
        
        \node[lbl] at (6.5, -.25) {Inputs};
        \node[lbl] at (7.5, -.25) {Outputs};
        
        \draw[>=stealth,->, thick] (input2) -- (output2);
    
    \end{tikzpicture}
    \caption[]{The multi-input heuristic clusters addresses of inputs jointly spent in the same transaction. It does not cluster addresses that are never co-spent with other addresses (such as \textsf{C} and \textsf{E}).}
    \label{fig:multi-input-problem}
\end{figure}

The current state of address clustering techniques available to researchers is sub-optimal in multiple ways.
The most common heuristic, \textit{multi-input}, groups addresses that are jointly used in inputs of a transaction \cite{reid2013,ron2013quantitative}. %
This heuristic is easy to apply, moderately effective in practice \cite{Harrigan2016}, and widely used.
However, it misses addresses that are never co-spent with other addresses (cf. \Cref{fig:multi-input-problem}).

Many of these addresses can be clustered using \textit{change address} heuristics: as coins in Bitcoin cannot be spent partially, transactions return the surplus value back to the sender.
Identifying the change output thus allows grouping the associated address with the inputs' addresses.
However, as the Bitcoin protocol does not explicitly distinguish between change and spend outputs, heuristics need to be used to identify them.

While the importance of change address identification and clustering has been demonstrated empirically and through simulation \cite{Meiklejohn2013, androulaki2013}, it remains difficult to assess how well it works in practice.
A major issue is that researchers currently lack ground truth data on change outputs to assess the accuracy of individual heuristics.
We are only aware of one prior study from 2015 that exploited weaknesses in a  lightweight client \cite{nick2015data}, which allowed to extract the addresses of \num{37585} wallets to assess four different clustering heuristics.
Blockchain intelligence companies might possess manually curated and refined data sets and clusterings, but their techniques and data aren't openly available to researchers (or only shared in limited form, e.g., \cite{harlev2018breaking,weber2019anti}).
As a result, analyses of new heuristics often fall short of quantifying their accuracy and resort to analyzing the resulting clusterings only (e.g., \cite{chang2018improving,zhang2020heuristic}).
Furthermore, clustering is applied inconsistently across studies: many  forgo change address clustering entirely (e.g., \cite{jourdan2018characterizing,schatzmann2020bitcoin,BlockSci,maesa2018data}), whereas some simply apply a single change heuristic (e.g., \cite{conti2018economic, parino2018analysis}).

Considering this state of affairs, our goals in this paper are to address the lack of ground truth data and assessment methods, develop new techniques to apply heuristics to predict change and use them to create improved clusterings.

\newpage
\subsubsection{Contributions, methods and findings.}
\begin{enumerate}
\item \textbf{A new ground truth method and dataset:} We put forward a procedure to select and filter transactions for which the change output has been revealed on the blockchain.
Our approach exploits that future transactions of users can reveal change outputs in past transactions.
We extract a set of \num{35.26} million transactions, carefully filtered down from \num{53} million candidate transactions, that can be used as ground truth for validation and prediction.
(\Cref{sec:ground-truth})
\item \textbf{Evaluating existing heuristics:} 
We've compiled and evaluate a set of 26 change address heuristics based on previous literature and community resources.
Most heuristics individually produce few false positives at low to medium true positive rates.  
We find that due to changes in the protocol and usage patterns, heuristics wax and wane in their effectiveness over time, showing the need to use multiple heuristics and combine them in an adaptive way rather than rely on a fixed algorithm.
(\Cref{sec:individual-heuristics})
\item \textbf{Improved prediction:} We use a random forest classifier to identify change outputs and compare it against a baseline: the majority vote of individual heuristics. While machine learning has been used to classify the type of entity behind a transaction (e.g., \cite{harlev2018breaking,jourdan2018characterizing,toyoda2018multi,lin2019evaluation,bartoletti2018data,hu2019characterizing,weber2019anti}), to the best of our knowledge our work is the first to apply it to change identification.
Our random forest model outperforms the vote, correctly detecting twice as many change outputs for low false positive rates.
(\Crefrange{sec:threshold}{sec:model-validation})
\item \textbf{Preventing cluster collapse:}
We find that a naive clustering of predicted change outputs leads to cluster collapse, despite using a high threshold to prevent false positives.
We then apply constraints to the union-find algorithm underlying our clustering to prevent cluster collapse stemming from frequent, repeated interaction between entities.
This prevents large-scale cluster collapse while still enhancing a majority of the involved clusters. (\Cref{sec:clustering})
\item \textbf{Assessing impact:} We assess the impact our enhanced clustering has on two exemplary applications: cash-out flows from darknet markets to exchanges and the velocity of bitcoins. We find that the results of such typical longitudinal analyses are off by at least \SIrange{11}{14}{\percent} if they don't fully account for clustering.
(\Cref{sec:applications})
\end{enumerate}

\subsubsection{Limitations.}

Our results in this paper are limited by the availability of \enquote{real} (i.e. manually collected and validated) ground truth.
As such, our analysis should be treated as a first step towards better understanding the feasibility of change address detection and clustering.
However, we do not expect our high-level insights to change significantly in the light of minor corrections to our ground truth data set.
We make our data set publicly available to allow other researchers to evaluate it using their own private ground truth or analysis techniques.

Our extraction mechanism relies on change outputs revealed by the multi-input heuristic.
This heuristic is effective in practice \cite{Harrigan2016} and widely used, but vulnerable to false positives from techniques like CoinJoin and PayJoin that are intentionally designed to break the heuristic (e.g., \cite{meiklejohn2015overlays,Moeser2017,CoinJoin,BIP78}).
While we take measures to detect CoinJoin transactions and pre-existing cluster collapse, some errors can remain.
Furthermore, entities that more effectively prevent address reuse are less likely to be included in our data set.

\section{Building a Ground Truth Data Set} \label{sec:ground-truth}

\subsubsection{Core assumption.}
We focus on the feasibility of detecting the change output in Bitcoin transactions with exactly two spendable outputs, by far the most common type of transaction as of June 2021 (\SI{75.8}{\percent} of all transactions, see \Cref{fig:transaction-overview}).
Our core assumption is that one of these outputs is a payment, and the other output receives the change.
We call this type of transaction a \textit{standard} transaction, as they are created by typical end-user wallet software.\footnote{Our definition is unrelated to the \texttt{isStandard} test in the Bitcoin reference implementation that checks whether a transaction uses common script types.}

For transactions with only one output there is no good indicator to directly and reliably determine whether the output belongs to the same user.
The transaction may correspond to a user sweeping the balance of their wallet, but the destination address may not be under the same user's control (e.g., it could be managed by a cryptocurrency exchange). %

Transactions with more than two outputs are less likely to originate from an ordinary  wallet.
They may belong to an exchange that batches payouts to multiple users, or correspond to a restructuring of their hot and cold wallets.
Our assumption that exactly one of the outputs receives change may not hold here.

\subsubsection{Method.}

\begin{figure}[!t]
    \centering
    \begin{minipage}[t]{.45\textwidth}
    \vspace{0pt}
    \begin{tikzpicture}[
        inout/.style={
            rectangle,
            semithick,
            draw,
            rounded corners=5pt,
            minimum width=.7cm,
            minimum height=.7cm,
            font=\sffamily
        },
        lbl/.style={
            font=\scriptsize\sffamily
        }
    ]
        \draw[] (0,0) rectangle ++(2,2);
        \draw[dotted] (1,0) -- ++ (0,2);
        \node[inout, fill=snblue!15] at (.5, 1.45) {A};
        \node[inout, fill=snblue!15] at (.5, .55) {B};
        \node[inout, fill=snblue!30] at (1.5, 1.45) (output) {C};
        \node[inout] at (1.5, .55) {D};
        
        \node[lbl] at (.5, -.25) {Inputs};
        \node[lbl] at (1.5, -.25) {Outputs};
        
        \draw[rounded corners=2pt,] (3,0) rectangle ++(2,2);
        \draw[dotted] (4,0) -- ++(0,2);
        \node[inout, fill=snblue!30] at (3.5, 1.45) (input) {C};
        \node[inout, fill=snblue!15] at (3.5, .55) {A};
        \node[inout] at (4.5, 1.45) {E};
        \node[inout] at (4.5, .55) {F};
        
        \node[lbl] at (3.5, -.25) {Inputs};
        \node[lbl] at (4.5, -.25) {Outputs};
        
        \draw[>=stealth,->, thick] (input) -- (output);
    \end{tikzpicture}
    \end{minipage}\hfill
    \begin{minipage}[t]{.52\textwidth}
    \vspace{0pt}
    \caption{The multi-input heuristic adds address \textsf{C} to the same cluster as addresses \textsf{A} and \textsf{B}, thereby revealing it as the change address of the first transaction.}
    \label{fig:change-revealed}
    \vspace{0pt}
    \end{minipage}
\end{figure}
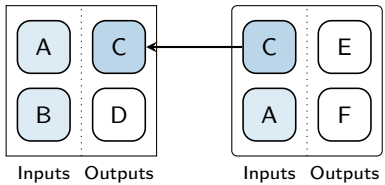

Our approach leverages that change outputs are sometimes revealed by the multi-input heuristic at a later point in time due to address reuse.
\Cref{fig:change-revealed} shows how such disclosure may unintentionally happen: a user spends coins at addresses $A$ and $B$, their wallet directs the change to a new address $C$.
Later, they spend the change at address $C$ along with other coins at address $A$.
At this point, the multi-input heuristic reveals that $A, B$ and $C$ belong to the same user, thus $C$ is the change address in the first transaction.
By identifying transactions that have their change revealed in this way, we can build a ground truth set of transactions with known change.

\subsubsection{Comparison to interactive collection.}

In contrast to prior deanonymization studies (e.g., \cite{Meiklejohn2013}) our primary interest is not in identifying address clusters of specific entities but to identify change outputs in their transactions.
To achieve this interactively, we would need to induce them to make a transaction to an address under our control.
This would likely yield inferior ground truth: %

\begin{itemize}
    \item Heterogeneous ground truth requires transactions from a \textit{variety of different use cases, entities and wallets}. We would only be able to directly interact with some types of intermediaries (such as exchanges). Our non-interactive method, instead, is not limited to a small set of intermediaries of our choosing.
    \item Interactive collection would be hard to \textit{scale} beyond a few hundred transactions, as we would have to individually engage with the intermediaries.
    Our non-interactive approach instead yields a data set of millions of transactions.
    \item Interactive collection cannot be done retroactively and is therefore limited to a short, current \textit{time frame}. The resulting data set wouldn't capture shifting patterns over different epochs of Bitcoin's history. Our non-interactive approach however can be applied to Bitcoin's entire history.
\end{itemize}

Our method has a few important limitations.
First, because we extract ground truth data non-interactively from the blockchain, we are not able to fully verify its correctness.
Second, our core assumption that exactly one of the outputs belongs to the user may not hold in every scenario.
For example, a user sending funds to an address under their control could lead to ambiguous or incorrect labeling of change outputs.
We take specific care to remove transactions likely to violate the core assumption in this way.
Similarly, there could be instances where none of the outputs is a change output.
As this would require a user to make a payment to two different entities using a perfectly matching set of inputs, we expect it to be rare.
Third, our ground truth set could be biased towards entities or wallet implementations that are more prone to reuse and merge addresses.

\subsection{Data collection and overview}

We use and build upon BlockSci v0.7 \cite{BlockSci}, an open-source blockchain analysis framework that provides fast access to blockchain data upon which we implement custom heuristics and extraction procedures.
We parse the Bitcoin blockchain until the end of June 2021 (block height \num{689256}) and create a \textit{base clustering} using the multi-input heuristic (where we heuristically exclude CoinJoin transactions).

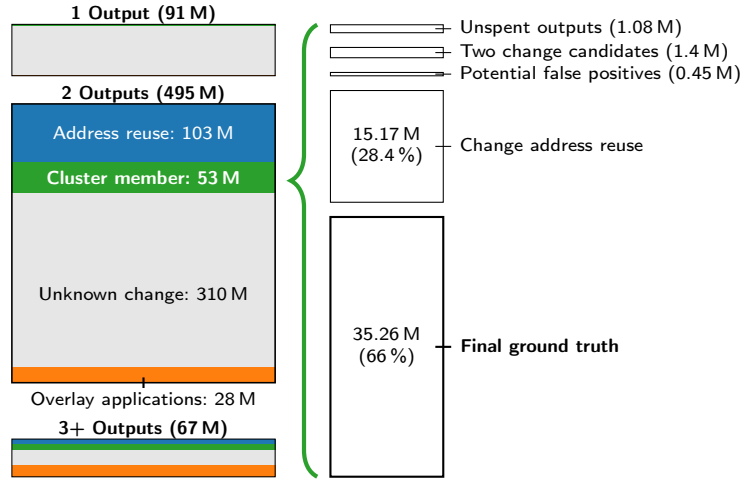
\begin{figure}
    \centering
    \begin{tikzpicture}[x=3.5cm, y=.075cm]
        
        \node[align=center,font=\scriptsize\sffamily\bfseries] at (.5, 70.32) {1 Output (91\,M)};
        \draw[fill=gray!20, draw=none] (0, 59.46) rectangle ++(1, 9.11);
        
        \draw[draw=none, fill=snblue] (0, 68.57) rectangle ++(1, -0.06);
        \draw[draw=none, fill=sngreen] (0, 68.51) rectangle ++(1, -0.22);
        \draw[draw=none, fill=snorange] (0, 59.46) rectangle ++(1, 0.0393);
        \draw (0, 59.46) rectangle ++(1, 9.11);
        
        \node[align=center,font=\scriptsize\sffamily\bfseries] at (.5, 56.21) {2 Outputs (495\,M)};
        
        \draw[fill=gray!20, draw=none] (0, 5) rectangle ++(1, 49.46);
        \draw[draw=none, fill=snblue] (0, 54.46) rectangle node {\scriptsize\sffamily\color{white} Address reuse: 103\,M} ++(1, -10.33);
        \draw[draw=none, fill=sngreen] (0, 44.13) rectangle node {\scriptsize\sffamily\bfseries\color{white} Cluster member: 53\,M} ++(1, -5.34);
        \draw[draw=none, fill=snorange] (0, 7.82) rectangle ++(1, -2.82);
        
        \node at (0.5, 20.5718) {\scriptsize\sffamily Unknown change: 310\,M};
        
        \draw[thick] ($ (0.5, 5) + (0, 2pt)$) -- ++(0, -4pt) node[below, inner sep=1pt] {\scriptsize\sffamily Overlay applications: 28\,M};
        \draw[semithick] (0, 5) rectangle ++(1, 49.46);
        
        \node[align=center,font=\scriptsize\sffamily\bfseries] at (.5, -3.25) {3+ Outputs (67\,M)};
        \draw[fill=gray!20, draw=none] (0, -10.6) rectangle ++(1, 5.6);
        \draw[draw=none, fill=snblue] (0, -5) rectangle ++(1, -0.907);
        \draw[draw=none, fill=sngreen] (0, -5.907) rectangle ++(1, -0.929);
        \draw[draw=none, fill=snorange] (0, -11.695) rectangle ++(1, 1.96);
        \draw (0, -11.695) rectangle ++(1, 6.695);

        \begin{scope}[
            shift={(1.21, 68.5)},
            x=1.5cm,
            y=.098cm,
            every node/.style = {font=\sffamily\scriptsize, align=center}]
            
            \draw (0, 0) rectangle (1, -1.08);
            \draw (0, -3.08) rectangle (1, -4.474909); %
            \draw (0, -6.474909) rectangle (1, -6.927850); %
            \draw (0, -8.927850) rectangle ++(1, -15.167451); %
            \draw[thick] (0, -26.095301) rectangle ++(1, -35.257428); %
            
            \draw [draw=sngreen, ultra thick, decorate, decoration={brace,amplitude=10pt, mirror, aspect=.345}] (-5pt, 0) --  ++(0, -61.35);
            
            \draw ($ (1, -.54) -(3pt, 0) $) -- ++ (6pt, 0) node[right] {Unspent outputs (1.08\,M)\Aq};
            
            \draw ($ (1, -3.7774545) -(3pt, 0) $) -- ++ (6pt, 0) node[right] {Two change candidates (1.4\,M)\Aq};
            
            \draw ($ (1, -6.7013795) -(3pt, 0) $) -- ++ (6pt, 0) node[right] {Potential false positives (0.45\,M)\Aq};
            
            \node at (.5, -16.5115755) {15.17\,M\\\scriptsize (28.4\,\%)};
            \draw ($ (1, -16.5115755) -(3pt, 0) $) -- ++ (6pt, 0) node[right] {Change address reuse\Aq};
            
            \node at (.5, -43.724015) {35.26\,M\\\scriptsize (66\,\%)};
            \draw[thick] ($ (1, -43.724015) -(3pt, 0) $) -- ++ (6pt, 0) node[right] {\textbf{Final ground truth\Aq}};
            
        \end{scope}

    \end{tikzpicture}
    \caption{Breakdown of different types of transactions in the Bitcoin blockchain until end of June 2021. Transactions with two outputs and change revealed through base cluster membership form the basis of our ground truth data, which we further refine to a final selection of \num{35.26} million transactions.}
    \label{fig:transaction-overview}
\end{figure}

\roundon{0}
As of June 2021, the blockchain contains \num{91.108655} million transactions with one output, \num{494.555377} million with two outputs, and \num{66.954143} million with three or more outputs (see \Cref{fig:transaction-overview}).
\roundoff
We divide the transactions into mutually exclusive categories.
Transactions with unspendable \opreturn{} outputs often signal the use of an overlay application that stores metadata in the blockchain \cite{Bartoletti2017}.
Such transactions may follow unique rules for their construction, potentially making  change detection unreliable.
Transactions directly reusing an input address have their change output trivially revealed and applying change heuristics is not necessary.
We thus focus on transactions where the change has been revealed by the multi-input heuristic and use them to construct our ground truth data set.
For the remaining transactions, i.e. those with yet unknown change, we will later predict their change output.

\subsection{Refining the candidate set of ground truth transactions}

Our candidate set of ground truth transactions consists of transactions with two outputs (ignoring overlay transactions) where no input address is reused for change and where at least one output is in the same base cluster as the inputs. This yields a total of  \num{53.41} million transactions.
We further filter them as follows (see \Cref{fig:transaction-overview} for a visual breakdown, and \Cref{app:gt-filtering} for additional details):

\roundon{2}
\begin{enumerate}
    \item We remove \num{1.084431} million transactions with unspent outputs, as our subsequent analyses rely upon the spending transactions being known.
    \item For \num{0.967601} million transactions both outputs are in the same base cluster, violating our core assumption. We remove these transactions. As some base clusters appear to be more likely to produce such transactions, we exclude transactions from base clusters where more than \SI{10}{\percent} of transactions exhibit this behavior. This removes \num{0.480845} million transactions in \num{9967} base clusters.
    \item We check our base clustering for preexisting cluster collapse, which could create false positives. We remove \num{0.366926} million transactions belonging to the Mt.Gox supercluster (cf. \cite{Harrigan2016}) as well as \num{0.087947} million transactions from one possible instance of cluster collapse detected using address tags from the website WalletExplorer.com.
    \item We find many instances where the change address did not appear in the inputs, but had been used before and was known to be the change at the time the transaction was created through multi-input clustering. For example, there are \num{5.77} million transactions originating from the gambling service \enquote{SatoshiDice} that use only a total of \num{50} change addresses, and \roundnum{2}{1.272124} million transactions from \enquote{LuckyB.it} that use a single change address. For such transactions, applying change address heuristics is never necessary. We remove \num{15.167451} million transactions where the change output was already known at the time the transaction was created.
\end{enumerate}
\roundoff

\subsection{Assessing the final set of ground truth transactions}

\subsubsection{Scale and time frame.}

\roundon{1}
Our final ground truth set of {\roundon{2}\num{35.257428}} million transactions makes up about  \SI{7.559995951268948}{\percent} of standard transactions and about \SI{5.402458796064023}{\percent} of all transactions.
These percentages are relatively stable over time.
\roundoff

\subsubsection{Variety of included clusters.}

\roundon{1}

\begin{figure}[!t]
    \centering
    \begin{subfigure}[t]{.49\textwidth}
        \centering
        \includegraphics[width=\columnwidth]{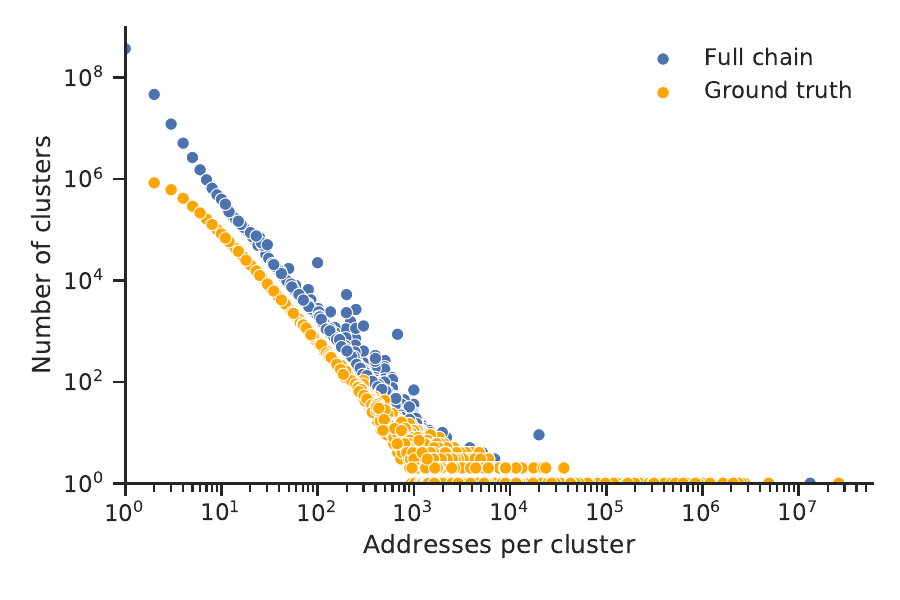}
        \caption{Number of base clusters represented in our ground truth by total address count.}
        \label{fig:cluster-sizes}    
    \end{subfigure}\hfill
    \begin{subfigure}[t]{.49\textwidth}
        \centering
        \includegraphics[width=\columnwidth]{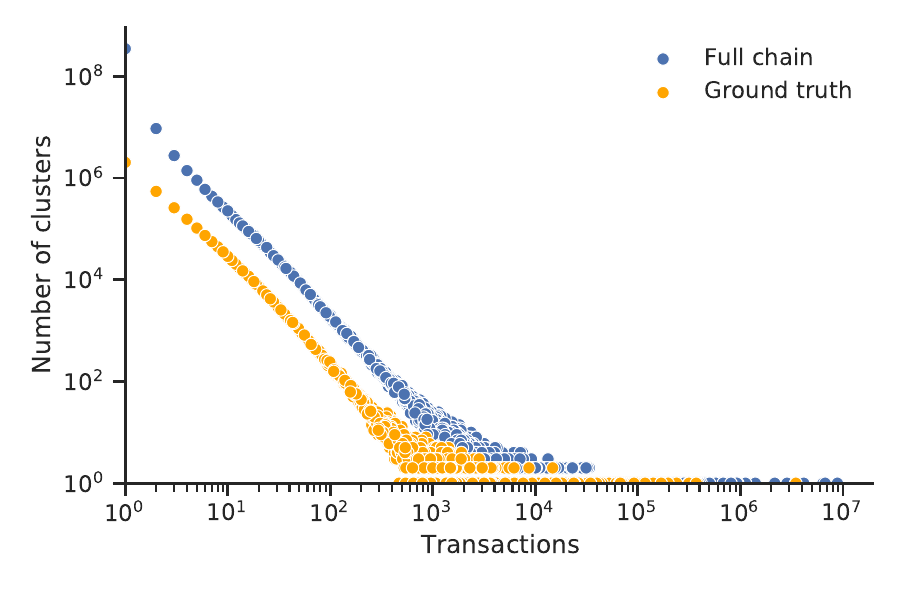}
        \caption{Number of transactions in ground truth and full blockchain per base cluster.}
        \label{fig:tx-counts}
    \end{subfigure}
    \caption{Address and transaction counts for base clusters in our ground truth.}
\end{figure}

Our ground truth includes transactions from \num{3.580558} million base clusters.
\Cref{fig:cluster-sizes} shows the distribution of address counts of base clusters that are represented with at least one transaction in our ground truth.
Our ground truth contains transactions from base clusters of all sizes, giving us confidence that it can be representative of the blockchain overall.

\Cref{fig:tx-counts} shows the number of transactions per base cluster included in the ground truth compared to the total number of transactions per cluster, showing an overall similar distribution.
The largest number of transactions from a single base cluster is \roundnum{2}{3.485257} million, which has \roundnum{2}{8.851845} million transactions in total.
We did not find a label for it on WalletExplorer.com.
The second highest number of transactions is \num{383519}, again from an unlabeled cluster.

\subsubsection{Transaction composition and use of protocol features.}

Compared to standard transactions with yet unknown change, our ground truth transactions have more inputs (\SI{38.92}{\percent} of transactions have three or more inputs, compared to  \SI{7.63}{\percent} for the  remaining transactions).
This is an expected artifact of our selection method, which relies on transactions with more than one input to reveal change outputs.
The share of transactions using SegWit serialization or allowing for fee bumping (RBF) is also higher in the set of remaining transactions.
In \Cref{app:ground-truth-characteristics} we provide additional details on the characteristics of our ground truth data.

\subsection{Data release}

We make our ground truth data set publicly available to allow other researchers to evaluate it using their own tools and techniques.\footnote{\url{https://github.com/maltemoeser/address-clustering-data}}
We believe that making this data public does not create significant new privacy risks: all information necessary to recreate the data set is already publicly available on the Bitcoin blockchain and our method---extracting change outputs revealed by the multi-input heuristic---is easy to reproduce with open-source tools like BlockSci.

\section{Predicting Change Outputs}\label{sec:ml-model}

The Bitcoin protocol does not explicitly distinguish between change and spend outputs.
However, wallets create change outputs automatically to return surplus value when users make payments.
This allows to guess the change using a variety of heuristics targeted at identifying specific wallet or user behavior.

In this paper we evaluate two general types of heuristics.
\textit{Universal} heuristics use characteristics of the transaction and change output to determine the change.
For example, the address type of a change output is likely to match the address types of the inputs, and rounded output values may indicate spend amounts.
\textit{Fingerprint} heuristics determine change based on matching characteristics of the transactions spending the outputs.
For example, if a transaction sets a positive locktime to prevent fee sniping \cite{Todd2014FeeSniping} and only one of the outputs is spent in a transaction with the same behavior, it is likely the change.
We are not aware of any prior work that has evaluated fingerprinting across the range of available protocol characteristics.
In total, we use \num{9} variants of universal heuristics and \num{17} variants of fingerprinting heuristics (cf. \Cref{tab:change-heuristics}).
To prevent cluster collapse, we explicitly encode our constraint that only one output can be the potential change: if both outputs are change candidates, none is returned by our heuristics.

\begin{table*}[]
    \centering
    \caption{Change heuristics proposed in the literature and used in this paper.}
    \label{tab:change-heuristics}
    \begin{tabularx}{\textwidth}{X @{~~} p{5cm} @{~} c}
    \toprule
    Heuristic & Notes and limitations & Used \\
    \midrule
    \textbf{Optimal change}: There should be no unnecessary inputs: if one output is smaller than any of the (2+) inputs, it is likely the change. \cite{nick2015data,BitcoinWikiPrivacy} & Only applies to transactions with 2+ inputs. We use two variants, one ignoring and one accounting for the fee. & $\checkmark$ \\
    \addlinespace
    \textbf{Address type}: The change likely uses the same address type as the inputs. \cite{BlockSci,BitcoinWikiPrivacy} & False positives possible due to protocol upgrades or obfuscation. & $\checkmark$ \\
    \addlinespace
    \textbf{Power of ten}: As purchase amounts may be rounded, and change amounts  depend on the input values and fee, it is more likely to have fewer trailing zeros. \cite{BlockSci,BitcoinWikiPrivacy} & We use six different variants, which are partially redundant. & $\checkmark$ \\
    \addlinespace
    \textbf{Shadow address}: Many clients automatically generate fresh change addresses, whereas spend addresses may be more easily reused. \cite{Meiklejohn2013,androulaki2013} & Modern wallets discourage reuse of receiving addresses. We do not use the heuristic as our ground truth is filtered based on address freshness. & \text{\sffamily x} \\
    \addlinespace
    \textbf{Consistent fingerprint}: The transaction spending a change output should share the same characteristics \cite{BitcoinWikiPrivacy,BlockchairPrivacyMeter}. We use 17 variants based on the following characteristics:
    \begin{itemize}[nosep]
        \item input/output counts and order
        \item version
        \item locktime
        \item serialization format (SegWit)
        \item replace-by-fee (RBF)
        \item transaction fee
        \item input coin age (zero-conf)
        \item address and script types
    \end{itemize}
    & False positives are possible when a wallet implementation or the protocol change. We only consider characteristics after they are available in the protocol. \Cref{app:characteristics} describes the characteristics we use in more detail. & $\checkmark$ \\
    \bottomrule
    \end{tabularx}
\end{table*}

\subsection{Assessing individual change heuristics}
\label{sec:individual-heuristics}

\roundon{2}

\begin{figure}[!t]
    \centering
    \includegraphics[width=\columnwidth]{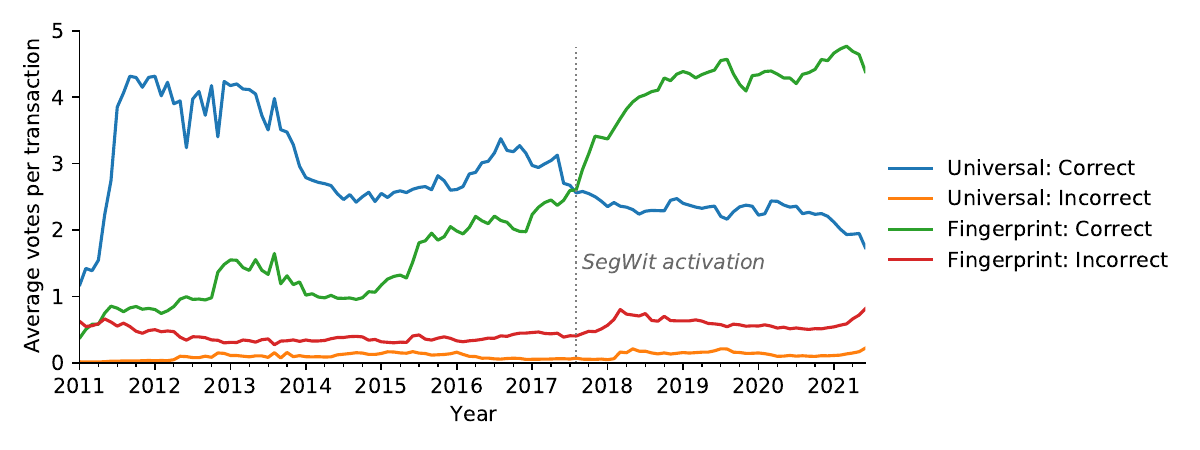}
    \caption{Average number of correct and uncorrect votes per transaction and type of heuristic in the ground truth data set, over time}
    \label{fig:heuristics-count}
\end{figure}

In a first step we assess each of the heuristics using our ground truth data set.
We find that most heuristics produce few false positives but often only apply to a small share of transactions (most heuristics have true positive rates between \SIrange{10}{30}{\percent}; detailed individual results are provided in \Cref{app:individual-heuristics-result}).
\Cref{fig:heuristics-count} shows the average number of correct and incorrect predictions per transaction over time, grouped by the type of heuristic.

We see three important trends: first, the universal heuristics drop over time, likely due to rounded values becoming less common.
Second, the consistent fingerprint heuristics see a steady uptick in the number of correct votes, enabled by the increasing variety of protocol features available in Bitcoin over time.
Finally, there's an uptick in both correct and incorrect fingerprint votes starting in late 2017, when wallet implementations started to switch to SegWit serialization and address formats (e.g., \cite{BitcoinCore16,CoinbaseSegWitFAQ}).

\roundoff

For \num{858582} transactions no heuristic returned a change output, we remove these from the subsequent analyses.
When we later predict change outputs for the remaining standard transactions, we will also exclude transactions where no heuristic determined a potential change output.

While most individual heuristics have high precision, they only cover a subset of transactions each.
Furthermore, some heuristics may be more applicable during certain epochs of Bitcoin's history than others.
Given the variety of heuristics available to us compared to previous studies (e.g., an evaluation of three change heuristics in \cite{nick2015data}), we now consider new ways of combining them.

\subsection{Threshold vote}
\label{sec:threshold}

\Cref{fig:heuristics-count} suggest that a majority of heuristics should generally identify the correct output.
However, the number of heuristics returning an potential output varies among transactions, and individual heuristics could be incorrect.
We thus compute a threshold vote: if at least $t$ more heuristics vote for output $a$ than for output $b$, then output $a$ is considered the change.
Increasing the threshold $t$ thus allows the analyst to require higher degrees of confidence and thereby lower the risk of cluster collapse.

\begin{figure}[!t]
\centering
\begin{subfigure}[t]{.49\textwidth}
  \centering
    \includegraphics[width=\columnwidth]{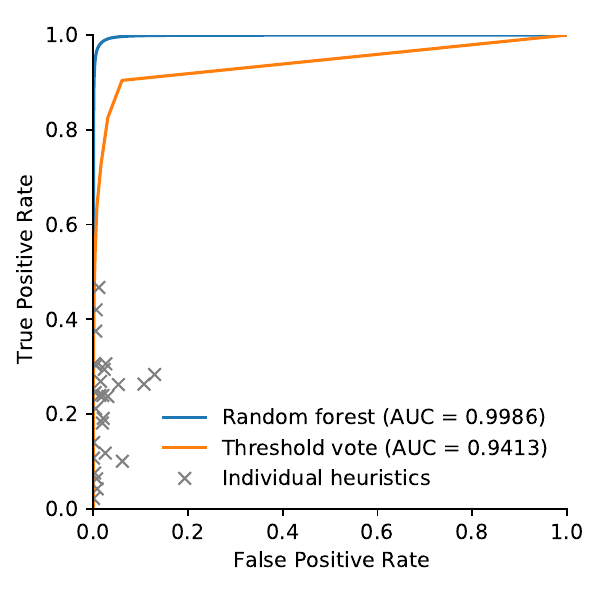}
    \caption{The random forest classifier outperforms the threshold vote and the individual heuristics.}
    \label{fig:threshold-vote-rf-roc}
\end{subfigure}\hfill
\begin{subfigure}[t]{.49\textwidth}
  \centering
    \includegraphics[width=\columnwidth]{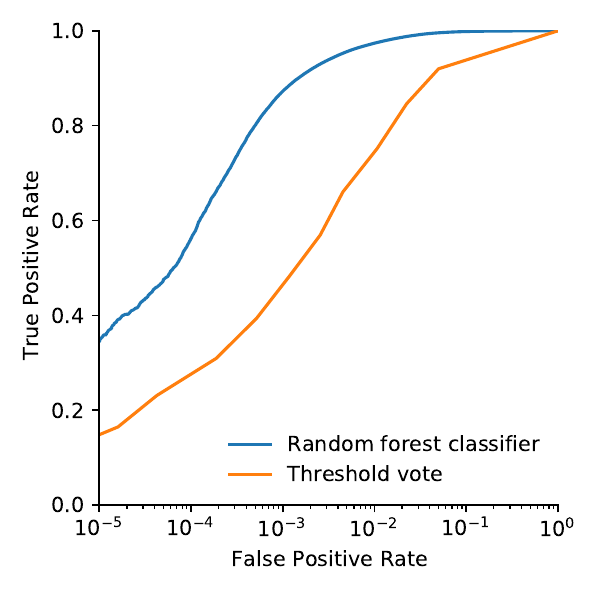}
    \captionof{figure}{The log scale highlights the difference between the classifiers for low false positive rates (on the same test set).}
    \label{fig:comparison-fpr}
\end{subfigure}
\caption{ROC curves for predicting change in the ground truth data set using the threshold vote and the random forest classifier, compared to individual heuristics.}
\end{figure}

We apply the threshold vote to our ground truth data set and plot the resulting ROC curve in \Cref{fig:threshold-vote-rf-roc} (for comparison, we also show the FPR and TPR of each individual heuristic).
We achieve an ROC AUC of \roundnum{2}{0.9413449324108675}, and, for example, a \SI[round-mode=places,round-precision=1]{37.0470}{\percent} true positive rate (TPR) below a false positive rate (FPR) of \SI{.1}{\percent} with a threshold of $t=7$.

Using a threshold vote may not be ideal as the individual heuristics have varying true positive and false positive rates, and some might be more or less reliable during different periods of Bitcoin's history.
Rather, a specific subset of heuristics may provide better classification accuracy.
Instead of manually trying different combinations, we opt to use a supervised learning classifier.

\subsection{Random forest classifier}

We decide to use a random forest classifier to predict a transaction's change output.
A random forest is an ensemble classifier that trains and aggregates the results of individual decision trees.
It is inherently well suited for our data set as it can divide it into homogeneous subsets, for example, based on protocol characteristics or time periods.
In an initial comparison of supervised classifiers on our data it also achieved the highest ROC AUC score (cf. \Cref{app:rf-tuning}).

We model an output-based binary classification problem, where 
every output is either a change (\num{1}) or spend (\num{0}) output.
An individual heuristic may produce one of three outcomes: vote for the output, against the output, or not be able to discern between the outputs.
We further add characteristics about each output and corresponding transaction that may allow the random forest to differentiate between distinct types of transactions, or wallets (see \Cref{app:rf-tuning} for details).

As we consider an analyst who works with a static snapshot of the blockchain, we randomly split our data set into \SI{80}{\percent} training and \SI{20}{\percent} test set.
We use the training set for hyperparameter tuning using 4-fold cross-validation, using the ROC AUC as our scoring metric.
To account for the fact that transactions in the same base cluster may be highly similar, we explicitly ensure that all outputs of a base cluster remain in the same set and fold.

Applying the random forest model (RF-1) to the test set, we achieve an AUC of \roundnum{4}{0.998622986868749} (\Cref{fig:threshold-vote-rf-roc}). %
The model is able to detect a higher share of outputs, especially at low false positive rates, compared to the threshold vote.

In \Cref{fig:comparison-fpr} we show the ROC curves of both the threshold vote and the random forest on the same test set, log-transforming the x-axis to highlight the important difference in low false positive rates (to prevent cluster collapse).
The random forest achieves much higher true positive rates at low false positive rates, meaning that it correctly identifies the change output in a larger number of transactions.
For example, if we target a false positive rate below \SI{0.1}{\percent}, the threshold vote achieves a TPR of around \SI{39}{\percent} at a FPR of \SI{0.06}{\percent}.
For the same FPR, the random forest achieves a TPR of \SI{82}{\percent}, more than twice as high.

We train a second random forest model (RF-2) without the fingerprint heuristics on transactions that contain predictions from the universal heuristics to later predict change in transactions with unspent outputs.
Using a similar evaluation strategy as for the full model, the ROC AUC of this model is \roundnum{4}{0.998071698922542}.

To ensure that the performance of our model is not dependent on the particular split and to determine its variance, we repeatedly split our ground truth data set into \SI{80}{\percent} training and \SI{20}{\percent} test set \num{20} times and train the random forest classifier using the previously determined hyperparameters.
The average ROC AUC score on the test sets is \roundnum{4}{0.9973833185025495} (SD = \roundnum{4}{0.001592319875743588}) for  RF-1, and \roundnum{4}{0.9965254404237829} (SD = \roundnum{4}{0.0036416234821124494}) for RF-2.

We note one caveat: because the base clustering is incomplete, grouping transactions by their base cluster may not fully prevent homogeneous transactions from the same entity to appear in both sets. %
Yet, some of the variability we see comes from unusual clusters that do not appear in the respective training sets.
Other researchers with private, more heterogeneous ground truth may be able to evaluate the degree to which this affects the overall performance of the model.

\subsection{Additional model validation}
\label{sec:model-validation}

We use two data sets to assess the performance of the random forest model trained on the entire ground truth data.
First, we use \num{16764} transactions identified by \textcite{Huang2018} as ransom payments related to the Locky and Cerber ransomware.
Those payments were identified through clustering, transaction graph analysis and known characteristics of the ransom amounts.
After removing non-standard transactions and those with revealed change output, we predict the change output for \num{11196} transactions and achieve an AUC of \roundnum{3}{0.9964282248572675}.

Our second data set is constructed using a GraphSense tagpack \cite{GraphSenseTagPacks} that contains \num{382} tags for addresses of \num{273} distinct entities (such as exchanges or gambling services) extracted from WalletExplorer.com.
We identify each associated cluster and then extract up to \num{1000} transactions occurring between the individual clusters, assuming that the output belonging to a different cluster is the spend output.
After removing transactions with no predictions as well as those with revealed change output,
we predict the change output for \num{268774} transactions and achieve an AUC of \roundnum{3}{0.9757453558776226}.

\section{Clustering Change Outputs}\label{sec:clustering}

We now use our random forest models to enhance the base clustering by clustering change outputs.
To this end, we predict the change outputs for \num{310} million standard transactions with yet unknown change.
We exclude \roundnum{1}{10.543720} million transactions where no individual heuristic identified a change output and use RF-2 for \roundnum{1}{19.294986} million transactions with unspent outputs.

To keep the likelihood of false positives low,  we use a conservative probability threshold of $p_{change} = 0.99$.\footnote{This corresponds to a false positive rate of \SI{0.044}{\percent} for RF-1. We use a threshold of \num{0.997} for RF-2 to match the FPR.}
\roundon{2}
This gives us \num{155.56} million change outputs (for \SI{50.24}{\percent} of  transactions).
We then enhance the base clustering by merging the base cluster of the inputs with the base cluster of the change address in the order that the transactions appear on the blockchain.
\roundoff

\subsection{Naive merging leads to cluster collapse}

Naively clustering the identified change outputs reduces \roundnum{1}{184.302672} million affected base clusters into \roundnum{1}{39.801817} million enhanced clusters.
However, it leads to severe cluster collapse: there is one large supercluster, containing the prior Mt.\ Gox supercluster, that contains \roundnum{1}{223.928805} million addresses (a \SI{1596}{\percent} increase) and \roundnum{1}{108.196018} million transactions (a \SI{2500}{\percent} increase).
Inspecting the \num{273} clusters labeled by the Graphsense tag pack, we find that \num{113} have been merged into the supercluster.

\subsection{Constraints prevent cluster collapse}

The majority of cluster merges involve address clusters from which only a single transaction originated.
Here, the impact of a single misclassification is low unless a sequence of such merges collapses multiple larger clusters.
At the same time, we observe a small number of merges that combine two large clusters.
Imagine two large exchanges whose users frequently interact with each other.
A single, misidentified change output could collapse their clusters.

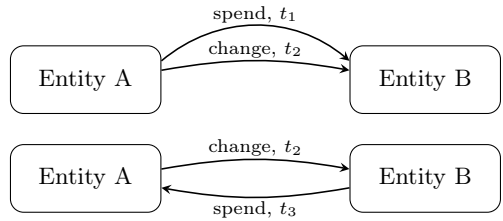
\begin{figure}[!t]
\centering
\begin{tikzpicture}[
    entity/.style = {
        draw,
        minimum height = .9cm,
        minimum width = 2cm,
        rounded corners=5pt
    },
    lbl/.style = {
        font = \scriptsize,
        inner sep = 1pt
    },
    >=stealth]
    \node[entity] (a) {Entity A\Aq};
    \node[entity, right = 2.5cm of a] (b) {Entity B\Aq};
    
    \draw[semithick, ->, bend left = 10] ([yshift=0.125 cm]a.east) to node [auto, lbl] {change, $t_2$} ([yshift=0.125 cm]b.west); 
    \draw[semithick, ->, bend left = 40] ([yshift=0.25 cm]a.east) to node [auto, lbl] {spend, $t_1$} ([yshift=0.25 cm]b.west); 
    
    \node[entity, below = .4cm of a] (a2) {Entity A\Aq};
    \node[entity, right = 2.5cm of a2] (b2) {Entity B\Aq};
    
    \draw[semithick, ->, bend left = 10] ([yshift=0.125 cm]a2.east) to node [auto, lbl] {change, $t_2$} ([yshift=0.125 cm]b2.west); 
    
    \draw[semithick, <-, bend right = 10] ([yshift=-0.125 cm]a2.east) to node [below, lbl] {spend, $t_3$} ([yshift=-0.125 cm]b2.west);

\end{tikzpicture}
\caption{In the pictured scenarios our constrained clustering prevents the merging of clusters A and B due to conflicting types of payments between them.} \label{fig:constrained-clustering}
\end{figure}

\subsubsection{Approach.}
We use this intuition to constrain which clusters we merge.
While change outputs predicted by our model should be clustered, we can use predicted spend outputs to prevent cluster merges: the input cluster should not be clustered into the cluster of the spend.
Given the probability $p_{i}$ returned by the random forest model for output $i$, we define two thresholds $p_{change}$ and $p_{spend}$ such that if $p_{i} > p_{change}$ the clusters should be merged (as before), but if $p_{i} < p_{spend}$ then the clusters should not be merged.
In many cases, these constraints will prevent the spend and change output of a transaction to end up in the same cluster (cf. \Cref{fig:constrained-clustering}).

This approach is comparable to that by \textcite{ermilov2017automatic} to use address tags in combination with a probabilistic model to reduce the number of conflicting tags in the final clustering.
However, public sources of address tags contain information on a limited number of intermediaries only.
Our approach, instead, potentially covers all clusters appearing in the \num{310} million standard transactions, including those that may be hard to interact with (and identify) manually.
Due to the size of our data set we only consider the binary case of preventing any potential conflict, accepting that we may prevent some valid merges in the process.

\roundon{2}
We implement a constrained union-find algorithm that prevents merging two clusters  related by a predicted spend output.
For every spend from cluster $c_m$ to cluster $c_n$, predicted with $p_i < p_{spend}$, we add a constraint to cluster $c_m$ that it must not be merged with cluster $c_n$.
Before merging two clusters, we the check the constraints of both clusters and skip the merge if it would violate them.

\subsubsection{Results.}
Using the same $p_{change} = 0.99$ and setting $p_{spend} = 0.01$, the constrained clustering prevents \num{413608} merges that would have violated constraints and retains \num{231340} more individual clusters than the unconstrained clustering.

We find that the constraints prevent the previously observed severe cluster collapse.
For example, the constrained clustering does not produce the large Mt.\ Gox supercluster: the cluster contains only \roundnum{1}{4.397689} million transactions (a \SI{6}{\percent} increase) and \roundnum{1}{14.511165} million addresses (a \SI{10}{\percent} increase).
Assessing the \num{273} labeled clusters, there are seven instances where two labeled clusters were merged.
We suspect that unusual types of payouts from these services might have triggered the collapse.

The largest cluster in the constrained clustering contains \roundnum{1}{20.379073} million transactions and \roundnum{1}{40.495271} million addresses.
Inspecting its composition, we find that it is the result of merging many small clusters (including \num{9421343} single-transaction clusters).

Overall, in at least \SI{90}{\percent} of merges the smaller cluster created at most one outgoing transaction, which highlights the usefulness of change address clustering to merge small clusters that are missed by multi-input clustering. %
The constrained clustering specifically prevents some of the largest merges observed in the naive clustering, thereby preventing cluster collapse.

\roundoff

\subsubsection{Varying thresholds.}

We chose conservative thresholds in order to reduce the possibility of cluster collapse.
At the same time, this means that fewer change outputs are being clustered than with lower thresholds.
To assess the impact of varying thresholds, we create two additional constrained clusterings, one with a threshold corresponding to a \SI{0.1}{\percent} FPR and one corresponding to a \SI{1}{\percent} FPR.
At \SI{0.1}{\percent}, the number of collapsed clusters identified by the Graphsense tag pack increases to \num{12}.
At \SI{1}{\percent}, however, there are already \num{60} instances of cluster collapse.
This highlights the importance of using conservative thresholds to prevent cluster collapse.

\section{Impact on Blockchain Analyses}\label{sec:applications}

Address clustering is a common preprocessing step before analyzing activity of entities on the blockchain.
Using different change heuristics (or none at all) thus affects the outcome of these analyses.

\subsection{Increased cashout flows from darknet markets to exchanges}

We evaluate the impact of our enhanced clustering on analysing payment flows from darknet markets to exchanges.
Such analyses are potentially relevant for cybercrime researchers, economists, regulators or law enforcement, highlighting the importance of address clustering for a variety of use cases.
To identify relevant intermediaries, we use address tags in the GraphSense tag pack for \num{117} exchanges and \num{15} darknet markets.

We extract the value of all outputs in transactions initiated by a darknet market that are sending bitcoins to an exchange, comparing the transaction volume calculated using our base clustering to that of our enhanced clustering.
The median increase in value sent across all \num{15} markets amounts to \SI{11.5}{\percent}.
The total amount of bitcoins flowing from the darknet markets to exchanges increases from BTC \num{823839} to BTC \num{937330} (a \SI{13.8}{\percent} increase).
We provide each individual market's increase in transaction volume in \Cref{app:darknet-exchanges}.

\subsection{Improved estimate of velocity}

We replicate the analysis of velocity conducted by \textcite{BlockSci}, an example for a longitudinal analysis of economic activity occurring on the Bitcoin blockchain.
For this analysis, clustering is used to remove self-payments of users (such as change outputs), which would artificially inflate estimates of economic activity.
The better and more complete our clustering, the more self-payments are removed and hence the lower the estimate will be.

Our refined clustering reduces their estimate of bitcoins moved per day between January 2017 to June 2021 by about \SI{11.9}{\percent}.
We notice that the magnitude is quite similar to the impact on cash-out flows.

\subsection{Comparison to the Meiklejohn et al. heuristic}

Finally, we compare our clustering to one created naively using the address reuse-based heuristic presented by \textcite{Meiklejohn2013}, which has subsequently been used in other studies (e.g., \cite{conti2018economic, parino2018analysis}).
While the authors highlight the need for manual intervention to prevent cluster collapse, this is likely infeasible for analysts without in-depth domain knowledge or the right set of tools.
The heuristic considers an output to be the change if its address has only been used a single time, based on common wallet behavior to not reuse change addresses.

Applying the heuristic to standard transactions with unknown change produces a supercluster containing \num{133.1} million transactions and \num{298.4} million addresses, with \num{177} tagged clusters ending up in the supercluster.
The probability of two addresses being clustered together increases by a factor of \num{40} compared to our constrained clustering. %
Looking at the individual predictions, the heuristic differs on \num{1.9} million transactions out of an overlapping \num{81.1} million.
The total pairwise difference in output values between those predictions amounts to BTC \num{4.1} million, or USD \num{38.7} billion, a significant difference in economic activity that might be misattributed due to clustering.

\section{Conclusion}\label{sec:conclusion}

Address clustering is an important cornerstone of many blockchain analyses.
In this paper, we've taken a first step towards building better models that allow analysts to identify change outputs in transactions, enabled by a new ground truth data set extracted from the Bitcoin blockchain.
Evaluating this data set, we find that for most transactions identifying the change address is feasible with high precision. %
Crucially, our work is the first to apply machine learning to the problem of change identification.
We find that our random forest model outperforms a baseline voting mechanism, detecting twice as many change outputs when targeting low false positive rates.
Turning to the subsequent clustering of change addresses, we've demonstrated that constraints based on our model's predictions can prevent cluster collapse.
Finally, we've explored the impact of our clustering on the outcome of economic analyses.
We hope that our work will encourage and enable further research into address clustering.

%% file: appendix.tex
\section{Additional details: Filtering the ground truth data set}\label{app:gt-filtering}

Selecting transactions with two outputs, no \opreturn{} outputs, where no input address has been directly reused in the outputs and where at least one output is in the same base cluster as the inputs yields a total of \roundnum{2}{53.412629} million transactions.
We first exclude \roundnum{2}{1.084431} million transactions with unspent outputs, as our subsequent analyses rely upon the spending transactions being known.

\paragraph{Transactions with two change candidates.}

Out of the \roundnum{2}{52.328198} million transactions with at least one change candidate, for \roundnum{2}{0.967601} million transactions \textit{both} outputs are in the same base cluster.
This can happen when a user transfers funds to an address in their own wallet, an online service restructures their funds, or cluster collapse leads to merging of both outputs' addresses.
In a first step, we exclude all transactions with two change candidates.

However, it is possible that there are yet unidentified transactions (due to an incomplete base clustering) where both outputs do belong to the same entity.
This should occur only in rare cases, but there may be specific intermediaries that create such transactions more frequently. %
We therefore exclude \textit{all} transactions from base clusters where more than \SI{10}{\percent} of transactions exhibit such behavior.
This removes an additional \num{480845} transactions in \num{9967} base clusters from our ground truth.

\paragraph{Potential false positives.}

A risk of using the base clustering to extract ground truth is that the multi-input heuristic could have produced false positives.
For example, if a user Alice makes a payment to merchant Bob and their wallet addresses are incorrectly clustered together, her spend output could appear to be the change.

To this end, we first remove \num{366926} transactions belonging to the Mt.\ Gox supercluster (cf. \cite{Harrigan2016}). %
Next, we spot-check our base clustering against the website \mbox{WalletExplorer.com}.
For the 100 largest base clusters in our ground truth we select 25 addresses at random and collect the tag (which is either explicitly named or pseudo-random) that WalletExplorer assigns to the address.
In five instances, the addresses yield multiple tags.
Four of these return only additional pseudo-random tags, which upon manual inspection we believe to be the result of a heuristic to not link addresses in transactions with large numbers of inputs.
Only one base cluster contains addresses with two different named tags: \enquote{LocalBitcoins.com-old} and \enquote{AnxPro.com}.
This could be a result of cluster collapse, or an instance of mislabeling on the side of WalletExplorer.
We remove the \num{87947} transactions from this base cluster from our ground truth.
Overall, this check gives us confidence that our base clustering does not already include widespread cluster collapse.

\roundon{2}
\paragraph{Change address reuse.}\label{sec:filter-freshness}

Our initial selection removed transactions where the change address appeared in an input of the transaction.
Yet, we find many instances where the change address did not appear in the inputs but had been seen before.
For example, a base cluster labeled by WalletExplorer as the gambling service \enquote{SatoshiDice}, contains \num{5.770524} million transactions that use only 50 different change addresses.
Similarly, there are \num{1.272124} million transactions from a base cluster tagged as \enquote{LuckyB.it} that all use a single change address.\footnote{\texttt{1NxaBCFQwejSZbQfWcYNwgqML5wWoE3rK4}}
In many of these cases, the change address could have already been revealed (before the transaction took place) through the multi-input heuristic.

If the change is known at the time the transaction is created, applying change heuristics is unnecessary.
In contrast, whenever a transaction uses a fresh address for change, it cannot possibly be revealed as the change at the time the transaction was created.
With this intuition, we remove transactions with change addresses that were not freshly generated if, at the time they were included in the blockchain, the change had already been revealed by the multi-input heuristic.
This removes a total of \num{15.167451} million transactions (\SI{90.8}{\percent} of transactions with reused change addresses).
\Cref{tab:fresh-non-fresh} provides an overview of whether the change and spend addresses are fresh in our ground truth data.

\roundoff

\begin{table}
    \centering
    \caption{Number of transactions (in million) in our ground truth data set with fresh or reused spend and change outputs.}
    \label{tab:fresh-non-fresh}
    \roundon{2}
    \begin{tabular}{l S S S}
    \toprule
         & \multicolumn{2}{c}{\textit{Spend}} \\
        \cmidrule(lr){2-3}
    \textit{Change} & {Reused} & {Fresh} & {Total} \\
    \midrule
    Reused & 0.730753 & 0.812165 & 1.542918 \\
    Fresh & 19.378533 & 14.335977 & 33.71451 \\
    \addlinespace
    Total & 20.109286 & 15.148142 & 35.257428 \\
    \bottomrule
    \end{tabular}
\end{table}

\section{Additional details: Assessing the final set of ground truth transactions}
\label{app:ground-truth-characteristics}

\subsubsection{Scale and time frame.}

\begin{figure}
    \centering
    \includegraphics[width=.75\columnwidth]{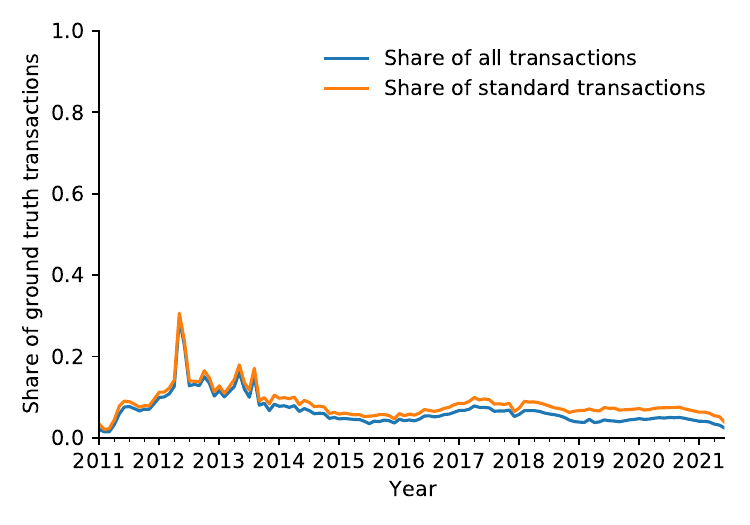}
    \caption{Share of ground truth transactions of all and standard transactions over time.}
    \label{fig:gt-share}
\end{figure}

\Cref{fig:gt-share} shows the share of ground truth transactions of all transactions and standard transactions over time. Overall, the distribution is relatively stable.

\subsubsection{Transaction composition and use of protocol features.}

\Cref{tab:gt-characteristics} compares characteristics of transactions in our ground truth data to those of standard transactions with yet unknown change, including the number of inputs as well as a number of important protocol features (an overview and description of the protocol characteristics used in this paper is available in \Cref{app:characteristics}).

\begin{table}[]
    \centering
    \caption{Comparison of transaction characteristics between ground truth transactions and transactions with 2 outputs for which change is unknown.}
    \label{tab:gt-characteristics}
    \roundon{2}
    \begin{tabular}{l S S}
        \toprule
        Characteristic & {Ground truth (\%)} & {Remaining (\%)} \\
        \midrule
        1 Input & 38.99156796122508 & 78.7611796007024 \\
2 Inputs & 22.088063258613193 & 13.609848007505846 \\
3+ Inputs & 38.92036878016173 & 7.62897239179175 \\

        \addlinespace
        Version = 1 & 79.83328222353599 & 80.68021282942037 \\
        Locktime > 0 & 25.25498456665642 & 24.59609683035776 \\
        RBF & 3.5686267302311445 & 6.221311481074558 \\
        SegWit & 18.29746344514977 & 27.09569756092249 \\
        \addlinespace
        $n$ (in million) & 35.257428 & 309.653054 \\
        \bottomrule
    \end{tabular}
\end{table}

\section{Protocol characteristics used for fingerprinting}
\label{app:characteristics}

\begin{itemize}
    \item Input/output count: the number of inputs and/or outputs may indicate a wallet software's behavior of creating transactions. While the number of inputs depends on the UTXOs available to the user, some commonly occurring patterns such as peeling chains (\cite{Meiklejohn2013,Moeser2013}) have consistent input and output counts.
    \item Version: BIP 68 \cite{BIP68} introduced relative timelocks for transactions, which requires transaction to set the transaction version to \texttt{2}.
    \item Locktime: Transactions can set a timelock such that they are valid only after the tip of the chain has passed a specific block height or timestamp. Some clients (e.g., Bitcoin Core) produce timelocked transactions by default to prevent fee-sniping \cite{Todd2014FeeSniping}.
    \item Replace-by-fee (RBF): Transactions opting into the replace-by-fee policy can be replaced by a similar transaction paying a higher fee \cite{BIP125}.
    \item SegWit: Segregated witness \cite{BIP141} is a protocol update that enabled storing the inputs' signatures outside of the transaction, thereby increasing available space in blocks. As the upgrade is backwards-comptabile, not all wallets produce SegWit transactions. A wallet might also be able to produce SegWit transaction, but may be required to use non-SegWit serialization if none of the inputs use SegWit. We call this behavior SegWit-conform.
    \item Ordered inputs/outputs: BIP 69 \cite{BIP69} defines non-binding rules (i.e. not enforced by the consensus mechanism) for lexicographically sorting inputs and outputs in a transaction. (A limitation of our implementation is that it does not compare the raw \texttt{scriptPubKey} in case the output values are equal, as they are not available in BlockSci).
    \item Zero-conf: Bitcoin user's are encouraged to wait for up to six confirmations (about an hour) before accepting a payment, as there is a risk that funds might be double-spent. A transaction spending inputs without any confirmations indicates willingness to accept the double-spending risk, which could be specific to certain intermediaries.
    \item Transaction fee: Bitcoin users pay transaction fees for their transactions to be included into the blockchain by miners. Some clients may pay the same exact fee (either absolute, or relative to the transaction's size) for every transaction.
    \item Multisignature: Multisignature scripts allow to specify a list of public keys and a threshold $m$ such that the redeemer must provide valid signatures for $m$ out of $n$ of these keys. They aren't typically used by normal end-user wallets.
    \item Address types: Bitcoin Core defines a number of standardized output scripts types including Pay-to-Pubkey-Hash (P2PKH), Pay-to-Script-Hash (P2SH) as well as their respective SegWit variants (P2WPKH and P2WSH). Often, a wallet consistently uses a specific address type.
    (Compared to the normal address type heuristic, the fingerprint checks for overlap with the address types of all inputs of the spending transaction).
\end{itemize}

\section{True and false positive rate of individual heuristics}
\label{app:individual-heuristics-result}

\begin{minipage}{\textwidth}
\begin{center}
    \centering
    \captionof{table}{True and false positive rates of each individual heuristic applied to transactions in our ground truth data set.}
    \label{tab:heuristics}
    \roundon{3}
    \begin{tabular}{l
        S[table-format=1.3]
        S[table-format=1.3]
        S[table-format=1.3]
        }
        \toprule
        & \multicolumn{2}{c}{Ground Truth} & {Remaining} \\
        \cmidrule(lr){2-3}\cmidrule(lr){4-4}
        {Heuristic} & {TPR} & {FPR} & {Coverage$^{*}$} \\
        \midrule
        {\textit{Universal heuristics}} \\
        {Optimal change} & 0.305982 & 0.026317 & 0.133055 \\
        {\tabitem incl. fee} & 0.2391 & 0.020221 & 0.095639 \\
        {Address type} & 0.236927 & 0.030544 & 0.368654 \\
        Power of ten & \\
        \tabitem $n = 2$ & 0.467048 & 0.011745 & 0.383023 \\
        \tabitem $n = 3$ & 0.419994 & 0.006165 & 0.310695 \\
        \tabitem $n = 4$ & 0.374834 & 0.005068 & 0.253267 \\
        \tabitem $n = 5$ & 0.302347 & 0.00557 & 0.172828 \\
        \tabitem $n = 6$ & 0.211132 & 0.004961 & 0.103664 \\
        \tabitem $n = 7$ & 0.106845 & 0.000501 & 0.048142 \\
        \addlinespace
        \multicolumn{2}{l}{\textit{Consistent fingerprint}}\\
        Output count & 0.283425 & 0.12945 & 0.444722 \\
        Input/output count & 0.263076 & 0.107035 & 0.568453 \\
        Version & 0.245298 & 0.003765 & 0.319966 \\ 
        Locktime & 0.306533 & 0.002925 & 0.362681 \\
        RBF & 0.075313 & 0.002698 & 0.114247 \\
        SegWit & 0.190708 & 0.020918 & 0.260257 \\
        SegWit-conform & 0.02142 & 0.000763 & 0.027523 \\
        Ordered ins/outs & 0.261993 & 0.052919& 0.442837 \\
        Zero-conf & 0.100237 & 0.06131 & 0.214499 \\
        Absolute fee & 0.117414 & 0.025319 & 0.305234 \\
        Relative fee & 0.041868 & 0.008099 & 0.203914 \\
        Multisignature & 0.139837 & 0.000551 & 0.153911 \\
        Address type & \\
        \tabitem P2PKH & 0.239226 & 0.01352 & 0.311971 \\
        \tabitem P2SH & 0.268645 & 0.015247 & 0.334274 \\
        \tabitem P2WPKH & 0.180882 & 0.019025 & 0.255838 \\
        \tabitem P2WSH & 0.06311 & 0.007178 & 0.08242 \\
        All address types & 0.294188 & 0.022907 & 0.391835 \\
        \bottomrule
        \multicolumn{4}{p{.55\columnwidth}}{\scriptsize$^{*}$Coverage denotes share of standard transactions with yet unidentified change where the heuristic returned exactly one output.}
    \end{tabular}
\end{center}
\end{minipage}

\section{Additional details: Random forest model}
\label{app:rf-tuning}

\subsubsection{Encoding.}

We transform our transaction-based predictions into an output-based binary classification problem.
Every output is either a change (\num{1}) or spend (\num{0}) output.
An individual heuristic may produce one of three outcomes: vote for the output, against the output, or not be able to discern between the outputs.
Due to the large size of the data set, we forgo one-hot encoding and instead use the following ordinal encoding for the heuristics:
\begin{itemize}
    \item[\textbf{1}] the heuristic votes for the output 
    \item[\textbf{0}] the heuristic votes neither for nor against the output 
    \item[\textbf{-1}] the heuristic votes against the output 
\end{itemize}

\subsubsection{Additional characteristics.}

We add the following characteristics about each output and corresponding transaction that may help the classifier differentiate between distinct types of transactions, or wallets.

\begin{itemize}
    \item Ratio of output's value to total transaction value
    \item Output index
    \item Total transaction value
    \item Transaction fee paid per byte
    \item Version number
    \item Non-zero locktime
    \item SegWit serialization
    \item Number of inputs
    \item Time of inclusion (as epochs of 1008 blocks, about one week)
\end{itemize}

\subsubsection{Baseline ROC AUC scores for different classifiers.}

Initial runs were done for a baseline comparison without hyperparameter tuning.
We note that our encoding may not be ideal for some classifiers, specifically for attributes that allow to subdivide behavior between different clients and epochs.
This is a major limitation of linear models and one of the primary reasons we choose a random forest model, as it is able to split the data set along those attributes.

\begin{itemize}
    \item Logistic regression (l2 penalty): \roundnum{4}{0.9932505213308463}
    \item Support Vector Machine (linear kernel): \roundnum{4}{0.9931019078568704}
    \item Adaboost: \roundnum{4}{0.9926415271932554}
    \item Random forest: \roundnum{4}{0.9982077979198866}
\end{itemize}

\subsubsection{Hyperparameter tuning for the random forest classifier.}
Our hyperparameter grid search returns the following parameters:
\begin{itemize}
    \item All heuristics / full model
    \begin{itemize}
        \item \texttt{max\_features: 7}
        \item \texttt{min\_samples\_leaf: 10}
        \item \texttt{min\_samples\_split: 20}
    \end{itemize}
    \item Universal heuristics only
    \begin{itemize}
        \item \texttt{max\_features: 6}
        \item \texttt{min\_samples\_leaf: 10}
        \item \texttt{min\_samples\_split: 20}
    \end{itemize}
\end{itemize}

\section{Change in transaction volume between darknet markets and exchanges}
\label{app:darknet-exchanges}

\begin{center}
    \captionof{table}{Change in outgoing transaction volumes of darknet markets (most of which were active between 2013 and 2016) to exchanges using the base clustering (before) and our enhanced clustering (after). }
    \label{tab:darknet-market-volumes}
    \begin{tabular}{
        l
        S[table-format=6.0]
        S[table-format=6.0]
        S[table-format=2.2]
    }
    \toprule
    & \multicolumn{2}{c}{Volume (BTC)}\\
    \cmidrule(lr){2-3}
    Tag name & {Before} & {After} & {Change (\%)} \\
    \midrule
    abraxasmarket & 21925 & 23368 & 6.58 \\
    agoramarket & 158360 & 170970 & 7.96 \\
    alphabaymarket & 35496 & 41573 & 17.12 \\
    babylonmarket & 222 & 283 & 27.13 \\
    blackbankmarket & 8292 & 9245 & 11.49 \\
    blueskymarket & 2520 & 3333 & 32.3 \\
    cannabisroadmarket & 6 & 7 & 25.15 \\
    doctordmarket & 224 & 277 & 23.92 \\
    evolutionmarket & 49891 & 84637 & 69.64 \\
    middleearthmarket & 11793 & 12021 & 1.93 \\
    nucleusmarket & 45265 & 47006 & 3.85 \\
    pandoraopenmarket & 8708 & 9461 & 8.64 \\
    sheepmarket & 12104 & 13309 & 9.96 \\
    silkroad2market & 47292 & 49559 & 4.79 \\
    silkroadmarket & 421741 & 472282 & 11.98 \\
    \addlinespace
    \textit{Total} & 823839 & 937330 & 13.78 \\
    \bottomrule
    \end{tabular}
\end{center}